\documentclass{appolb}
\usepackage{graphicx}
\usepackage{amssymb,amsmath,amsfonts}
\usepackage[utf8]{inputenc}
\usepackage[normalem]{ulem}
\usepackage{color}
\usepackage{xspace}

\newcommand{\ud}{\mathrm{d}}

\newcommand{\mub}{\ensuremath{\mu_{B}}\xspace}
\newcommand{\dndy}{\ensuremath{\ud N/\ud y}\xspace}

\newcommand{\jpsi}{\ensuremath{\mathrm{J}/\psi}\xspace}

\renewcommand\sout{\bgroup \color{blue} \ULdepth=-.5ex \ULset}

\begin{document}

\title{Hadron yields in central nucleus-nucleus collisions, the statistical hadronization model and the QCD phase diagram}

\author{Anton Andronic
  \address{Institut f\"ur Kernphysik, Universit\"at M\"unster, 48149  M\"unster, Germany}
\\
Peter Braun-Munzinger
\address{Research Division and EMMI, GSI Helmholtzzentrum f\"ur Schwerionenforschung, 64291 Darmstadt, Germany}
\address{Physikalisches Institut, Universit\"at Heidelberg, 69120 Heidelberg, Germany}
\address{Institute of Particle Physics and Key Laboratory of Quark and Lepton Physics (MOE), Central China Normal University, Wuhan 430079, China}
\\
Krzysztof  Redlich
\address{University of Wroc\l aw, Institute of Theoretical Physics, 50-204 Wroc\l aw, Poland}
\\
Johanna  Stachel
\address{Physikalisches Institut, University of Heidelberg, 69120 Heidelberg, Germany}
}

\maketitle 

\begin{abstract}

The description of hadron production in relativistic heavy-ion collisions in the statistical hadronization model is very good, over a broad range of collision energy. We outline this both for the light (u, d, s) and heavy (charm) quarks and discuss the connection it brings to the phase diagram of QCD.

\end{abstract}

\section{Introduction} \label{sect:intro}

If one compresses or heats nuclear matter to higher densities and/or high temperatures one expects \cite{Itoh:1970uw,Collins:1974ky,Cabibbo:1975ig,Chapline:1976gy} that quarks are no longer confined but can move over distances significantly larger than the size of the nucleon.
Such a deconfined state of matter, the Quark-Gluon Plasma (QGP) \cite{Shuryak:1978ij}, is likely to have existed in the Early Universe within the first (about 10) microseconds after its creation in the Big Bang \cite{Boyanovsky:2006bf} and is studied experimentally and theoretically via collisions of nuclei at high energies \cite{Busza:2018rrf,Braun-Munzinger:2015hba}.
One stage in the complex dynamics of the system produced in heavy-ion collisions is that of the chemical freeze-out, at which the abundance of hadron species is fixed (frozen), addressed phenomenologically within the statistical hadronization model (SHM) \cite{Andronic:2017pug}.
The value of the crossover temperature $T_c$ at vanishing \mub is currently calculated in Lattice QCD (LQCD) to be 156.5$\pm 1.5$ MeV \cite{Bazavov:2018mes} and 158.0$\pm 0.6$ MeV \cite{Borsanyi:2020fev}.
Recent LQCD results also quantify the small decrease of $T_c$ with increasing \mub as long as $\mub \lesssim 300$ MeV \cite{Bonati:2018nut,Bazavov:2018mes,Borsanyi:2020fev}. Within this parameter range the transition is still of crossover type \cite{Aoki:2006we}.

One of the consequences of confinement in QCD is that physical observables require a representation in terms of hadronic states. Indeed, as has been noted in the context of QCD thermodynamics (see, e.g., \cite{Bazavov:2017dus} and refs. therein) the corresponding partition function $Z$ can be very well approximated within the framework of the hadron resonance gas, as long as the temperature stays below $T_c$.

The grand canonical partition function for specie (hadron) $i$ is: 
\begin{equation}
\ln Z_i ={{Vg_i}\over {2\pi^2}}\int_0^\infty \pm p^2\ud p \ln [1\pm  \exp (-(E_i-\mu_i)/T)]
\end{equation}
with $+$ for fermions and $-$ for bosons, where $g_i=(2J_i+1)$ is the spin degeneracy factor, $T$ is the temperature, $E_i =\sqrt {p^2+m_i^2}$ the total energy;
$\mu_i = \mu_B B_i+\mu_{I_3} I_{3i}+\mu_S S_i+\mu_C C_i$ are the chemical potentials ensure conservation (on average) of baryon, isospin, strangeness and charmness quantum numbers.  Three initial conditions help fixing ($I_{3i},\mu_S,\mu_C$):
i) isospin stopping identical to baryon stopping: $I_{3}^{tot}/\sum_i n_i I_{3i}= N_B^{tot}/\sum_i n_i B_i$, with $I_{3}^{tot}$, $N_{B}^{tot}$ isospin and baryon numbers of the system (proportional to $\mu_B/931$, with $\mu_B$ reflecting baryon stopping in the collision);
ii) vanishing net initial strangeness: $\sum_i n_i S_i = 0$; iii) vanishing net initial charmness: $\sum_i n_i C_i = 0$.

On needs for the calculations the knowledge of the complete hadron spectrum and the default constitutes what is listed by PDG \cite{Zyla:2020zbs}; the presence of resonances corresponds to attractive interactions among hadrons.
Traditionally, repulsive interactions are modelled with an 'excluded volume' prescription \cite{Rischke:1991ke}. 
For weak repulsion, implying excluded volume radii $r_0 \le 0.3$ fm, the effect of the correction is a decrease of particle densities, while the important thermal parameters $T$ and \mub are little affected. Strong repulsion cannot be modelled that way: significantly larger $r_0$ values lead to, among others, unphysical (superluminous) equations of state, in contra-distinction to results from LQCD.
Other approaches, like temperature-dependent resonance widths \cite{Vovchenko:2018fmh} were recently proposed, but lack full consistency. 
A consistent approach is the implementation employing the S-matrix formulation of statistical mechanics with measured pion-nucleon interactions including, importantly, also non-resonant components \cite{Andronic:2018qqt}. In this approach, currently implemented only for $\mub\simeq 0$ (and here for the non-strange sector), the effect of multi-pion-nucleon interactions is estimated using LQCD.

\section{Statistical hadronization of light quarks}

\begin{figure}[hbt]
\includegraphics[width=.49\textwidth]{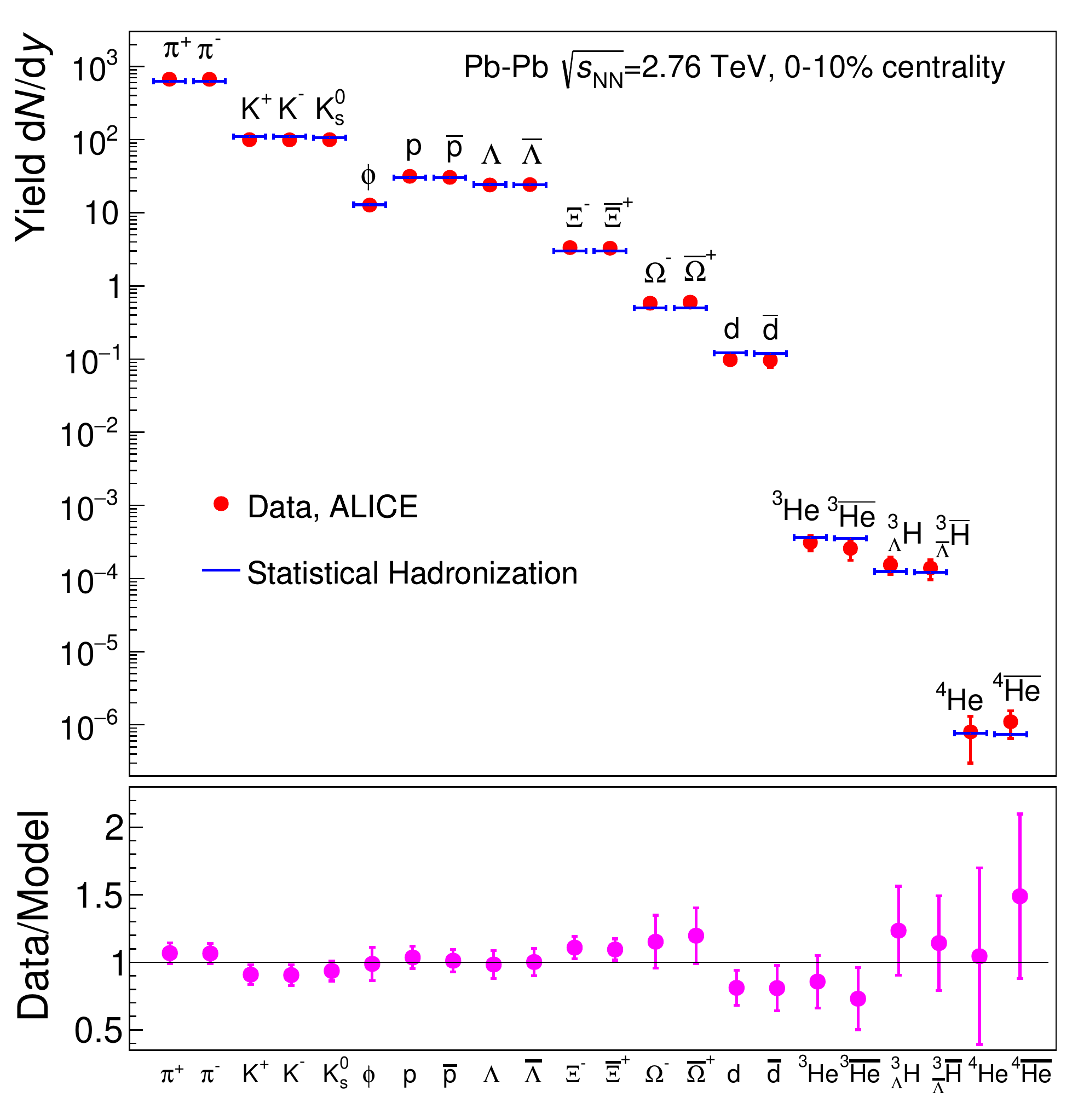}
\includegraphics[width=.49\textwidth]{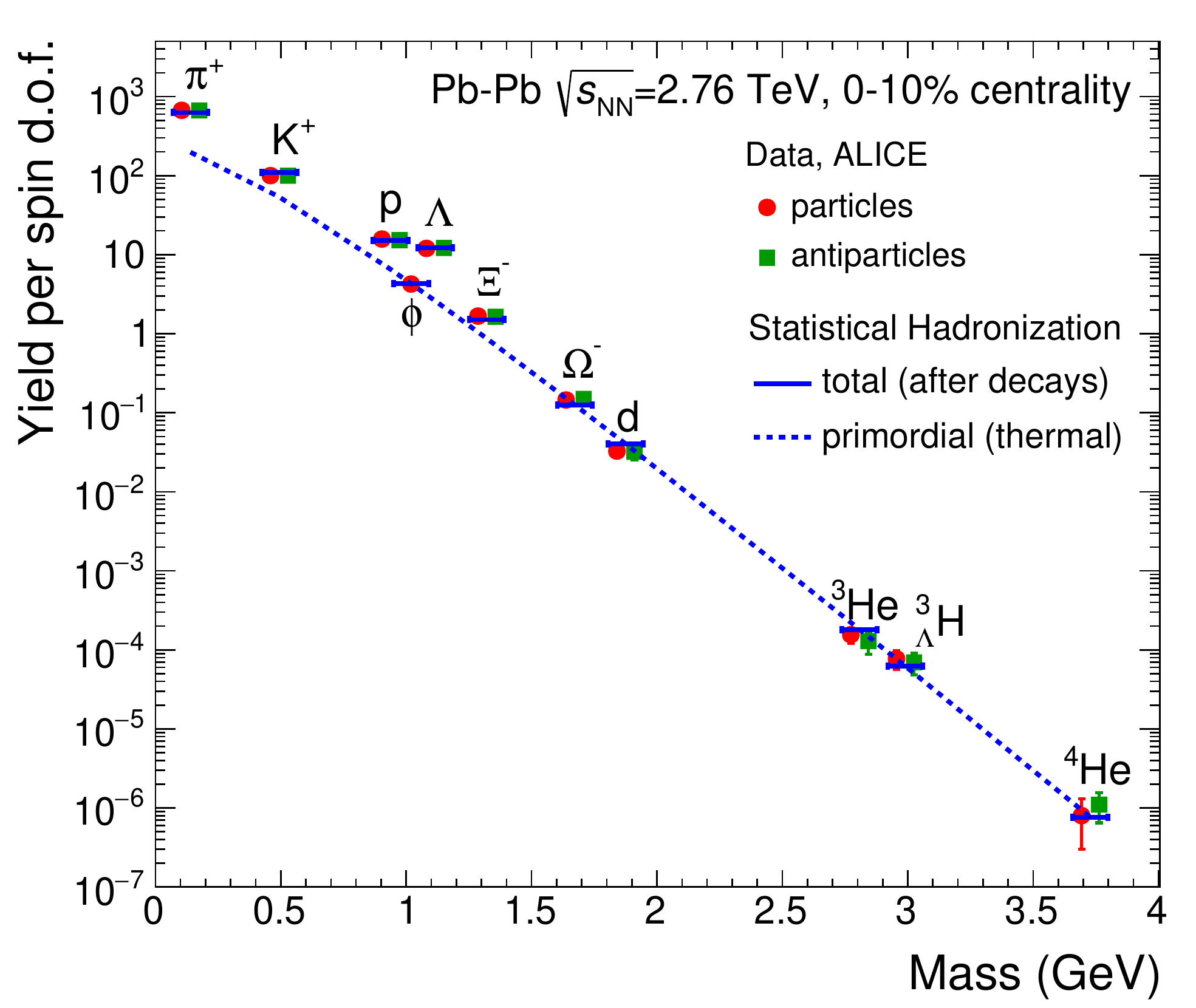}
\caption{Left: Hadron yields $\ud N/\ud y$ measured in central Pb--Pb collisions at the LHC and the best fit with SHM. The lower panel shows the ratio of data and model with uncertainties (statistical and systematic added in quadrature) of the data.
Right: Mass dependence of hadron yields divided by the spin degeneracy factor ($2J+1$). For SHM, plotted are the ``total'' yields, including all contributions from high-mass resonances (for the $\Lambda$ hyperon, the contribution from the electromagnetic decay $\Sigma^0\rightarrow\Lambda\gamma$, which cannot be resolved experimentally, is also included), and the (``primordial'') yields prior to strong and electromagnetic decays.
}
\label{fig:Fit}
\end{figure}

In practice, $T_{CF}$, $\mub$, and $V$, the parameters at chemical freeze-out are determined from a fit to the experimental data. 
For the most-central (0-10\%) Pb--Pb collisions at the LHC, the best description of the ALICE data (see \cite{Acharya:2017bso} and ref. therein) on yields of particles in one unit of rapidity at midrapidity, is obtained with $T_{CF}=156.6\pm 1.7$ MeV, $\mu_B=0.7\pm 3.8$ MeV, and $V=4175\pm 380$ fm$^3$ (corresponding to a slice of one unit of rapidity, centered at mid-rapidity) \cite{Andronic:2018qqt}, shown in Fig.~\ref{fig:Fit}.
The standard deviations quoted here are exclusively due to experimental uncertainties and do not reflect the systematic uncertainties connected with the model implementation.

Very good agreement is obtained between the measured particle yields and SHM over nine orders of magnitude in abundance values and encompasses strange and non-strange mesons, baryons including strange and multiply-strange hyperons as well as light nuclei and hypernuclei and their anti-particles.
The initially-observed overprediction of data by the model for proton and antiproton yields (a deviation of 2.7$\sigma$) is entirely accounted for via the S-matrix treatment of the interactions \cite{Andronic:2018qqt} included here (for consistency the excluded-volume correction is not applied anymore). It was recently shown that the addition (compared to what is listed by PDG \cite{Zyla:2020zbs}) of about 500 new states predicted by LQCD and the quark model does lead to a deterioration of the fit, while no change is observed when the S-matrix treatment is employed \cite{Andronic:2020iyg}. 

The thermal origin of all particles including light nuclei and anti-nuclei is particularly transparent when inspecting the dependence of their yields with particle mass, shown in the right panel of Fig.~\ref{fig:Fit}.
We note that the yields of the measured lightest mesons and baryons, ($\pi,K,p,\Lambda$) are substantially increased relative to their primordial thermal production by the resonance decay contributions (for pions, e.g., the decay contribution amounts to 70\% of the total yield).
For the subset of light nuclei, the SHM predictions are, however, not affected by resonance decays.
For these nuclei, a small variation in temperature leads to a large variation of the yield, resulting in a relatively precise determination of the freeze-out temperature $T_{nuclei} = 159 \pm 5$ MeV, well consistent with the value of $T_{CF}$ extracted above.

The rapidity densities of light (anti)-nuclei and hypernuclei were actually predicted \cite{Andronic:2010qu}, based on the systematics of hadron production at lower energies. It is nevertheless remarkable that such loosely bound objects (the deuteron binding energy is 2.2 MeV, much less than $T_{CF} \approx T_c  \approx 157$ MeV) are produced with temperatures very close to that of the phase boundary at LHC energy, implying any further evolution of the fireball has to be close to isentropic.
The detailed production mechanism for loosely bound states remains an open question (see recent review \cite{Braun-Munzinger:2018hat}). 
  One possibility, considered  already long ago \cite{Chapline:1978kg}, is that such objects, at QGP hadronization, are produced as compact, colorless droplets of quark matter with quantum numbers of the final state hadrons.

\begin{figure}[htb]
  \includegraphics[width=.46\textwidth]{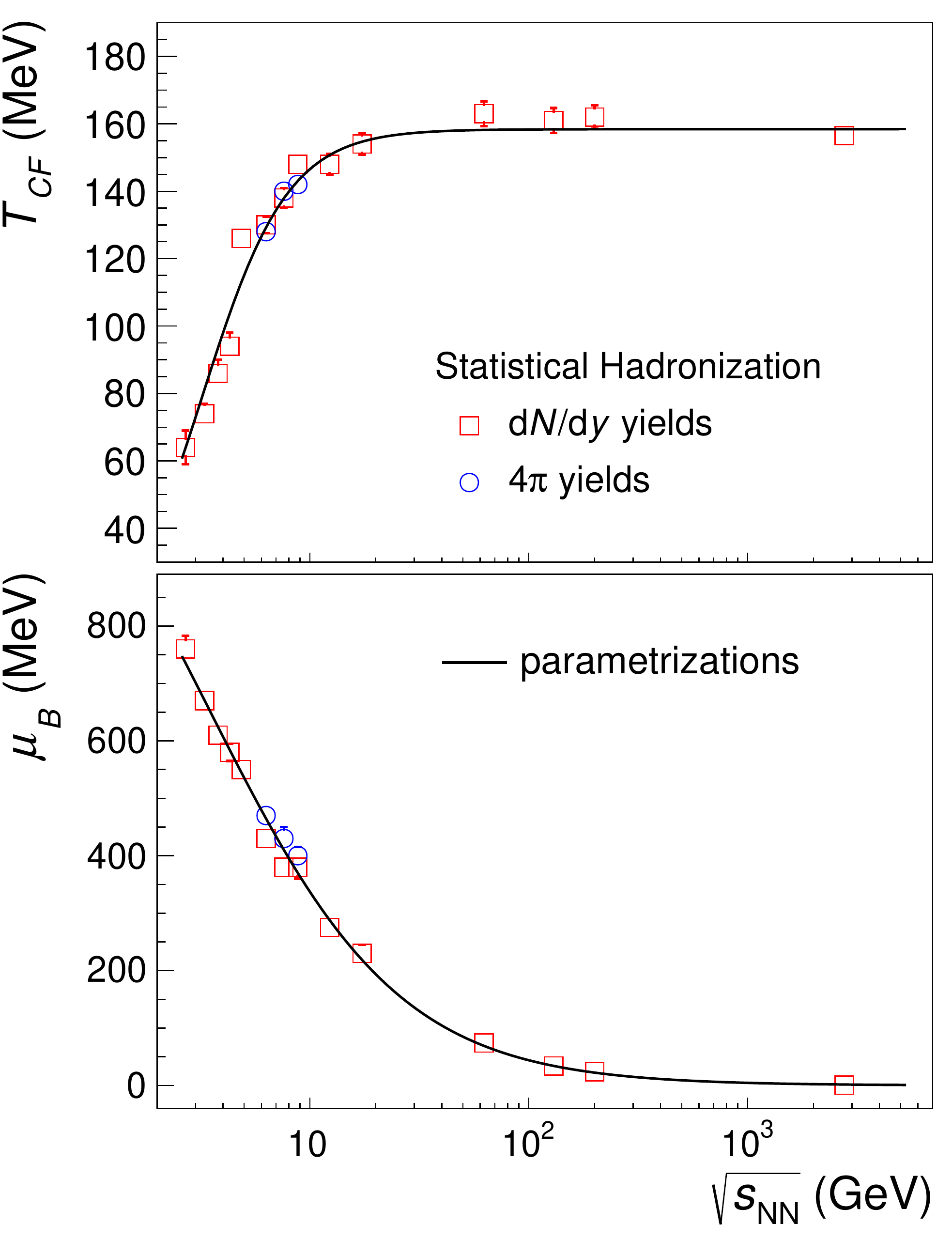}
\includegraphics[width=.52\textwidth]{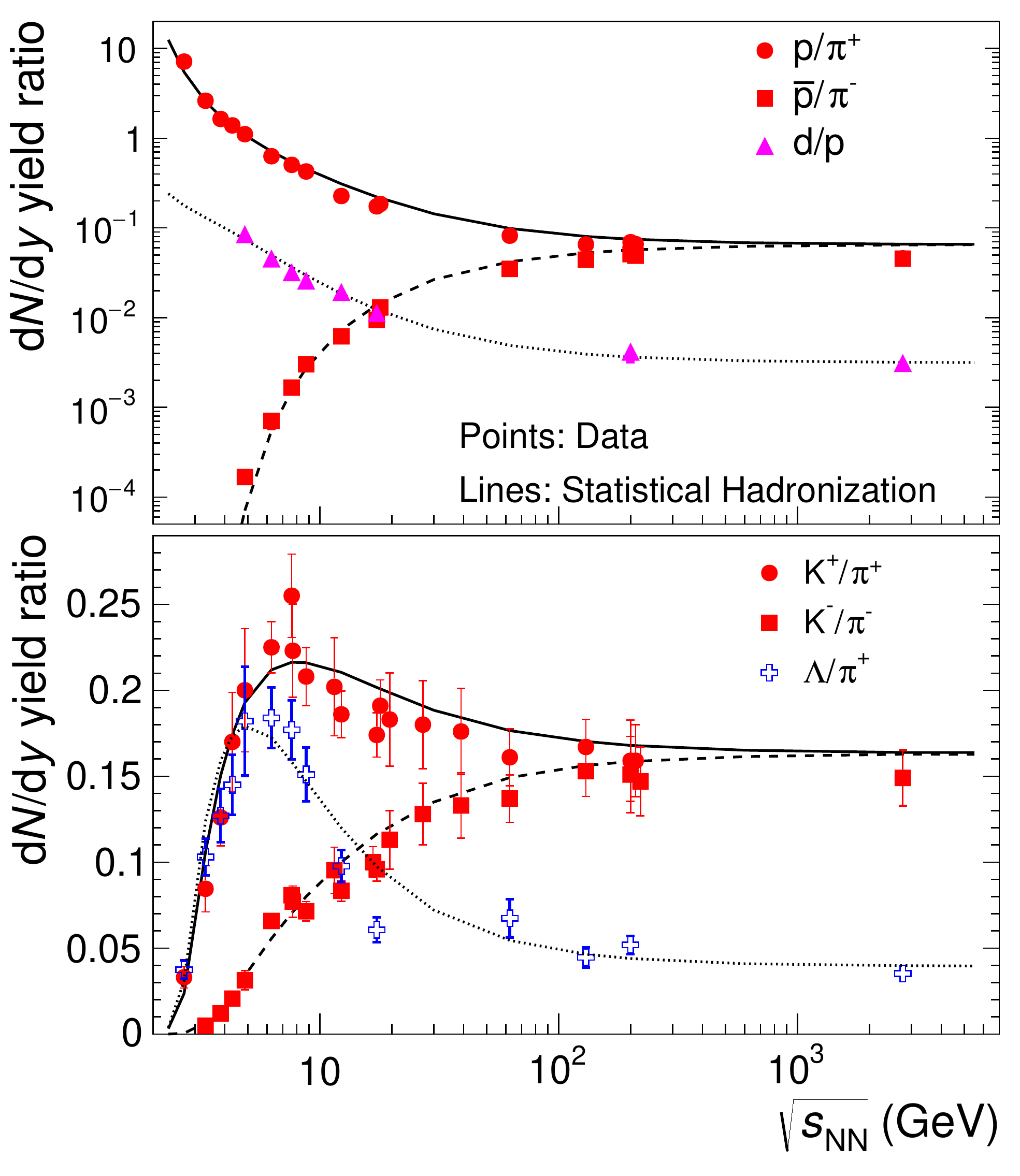}
\caption{Left: Energy dependence of chemical freeze-out parameters $T_{CF}$
  and \mub. The results are obtained from the SHM analysis of hadron yields (at midrapidity, \dndy, and in full phase space,  $4\pi$) for central collisions at different energies. Right: Collision energy dependence of the relative abundance of several hadron species (the data are compiled in \cite{Andronic:2014zha,Adamczyk:2017iwn}).}
\label{fig:edep}
\end{figure}

The thermal nature of particle production in ultra-relativistic nuclear collisions has been experimentally verified not only at LHC energy, but also at the lower energies of the RHIC, SPS and AGS accelerators. The essential difference is that, at these lower energies, the matter-antimatter symmetry observed at the LHC is lifted, implying non-vanishing values of the chemical potentials. Furthermore, in central collisions at energies below $\sqrt{s_{\rm NN}} \approx 6 $ GeV the cross section for the production of strange hadrons decreases rapidly, with the result that the average strange hadron yields per collision can be significantly below unity. In this situation, one needs to implement exact strangeness conservation, applying the canonical ensemble for the conservation laws \cite{Hagedorn:1984uy,Hamieh:2000tk}. Similar considerations apply for the description of particle yields in peripheral nuclear and elementary collisions.
A consequence of exact strangeness conservation is the suppression of strange particle yields when going from central to peripheral nucleus-nucleus collisions or from high multiplicity to low multiplicity events in proton-proton or proton-nucleus collisions \cite{ALICE:2017jyt,Cleymans:2020fsc}.

While \mub decreases smoothly with increasing energy, the dependence of $T_{CF}$ on energy exhibits a striking feature which is illustrated in Fig.~\ref{fig:edep}: $T_{CF}$ increases with increasing energy (decreasing \mub) from about 50 MeV to about 158 MeV, where it exhibits a saturation for $\sqrt{s_{\rm NN}} > 20$ GeV.  The slight increase of this value compared to $T_{CF} = 156.6$ MeV obtained at LHC energy is due to the inclusion of points from data at RHIC energies, the details of this small difference are currently not fully understood.
The saturation of $T_{CF}$ observed in Fig.~\ref{fig:edep} lends support to the earlier proposal \cite{BraunMunzinger:1998cg,Stock:1999hm,BraunMunzinger:2003zz} that, at least at high energies, the chemical freeze-out temperature is very close to the QCD hadronization temperature \cite{Andronic:2008gu}, implying a direct connection between data from relativistic nuclear collisions and the QCD phase boundary.
This is in accord with the earlier prediction, already more than 50 years ago, by Hagedorn \cite{Hagedorn:1965st} that hadronic matter cannot be heated beyond this limit. 
The parametrizations shown in Fig.~\ref{fig:edep} are:
$T_{CF}={T_{CF}^{lim}}/(1+\exp(2.60-\ln(\sqrt{s_{\rm NN}})/0.45))$,
$\mu_B ={a}/(1+0.288\sqrt{s_{\rm NN}})$, with $\sqrt{s_{\rm NN}}$ in GeV and 
 the 'limiting temperature' $T_{CF}^{lim}=158.4\pm 1.4$ MeV and $a=1307.5$ MeV. 

To illustrate how well the thermal description of particle production in central nuclear collisions works we show, in Fig.~\ref{fig:edep} (right), the energy dependence (excitation function) of the relative abundance of several hadron species along with the prediction using the SHM and the parametrized evolution of the parameters.
 In particular, the maxima  (occuring at slightly different c.m. energies) in the $K^+/\pi^+$ and $\Lambda/\pi^+$ ratios are naturally explained \cite{Andronic:2008gu} as the interplay between the energy dependence of $T_{CF}$ and $\mu_B$ and the consequence of strangeness conservation.

\begin{figure}[htb]
\begin{tabular}{cc}  \begin{minipage}{.48\textwidth}
 \hspace{-.4cm}   \includegraphics[width=1.04\textwidth]{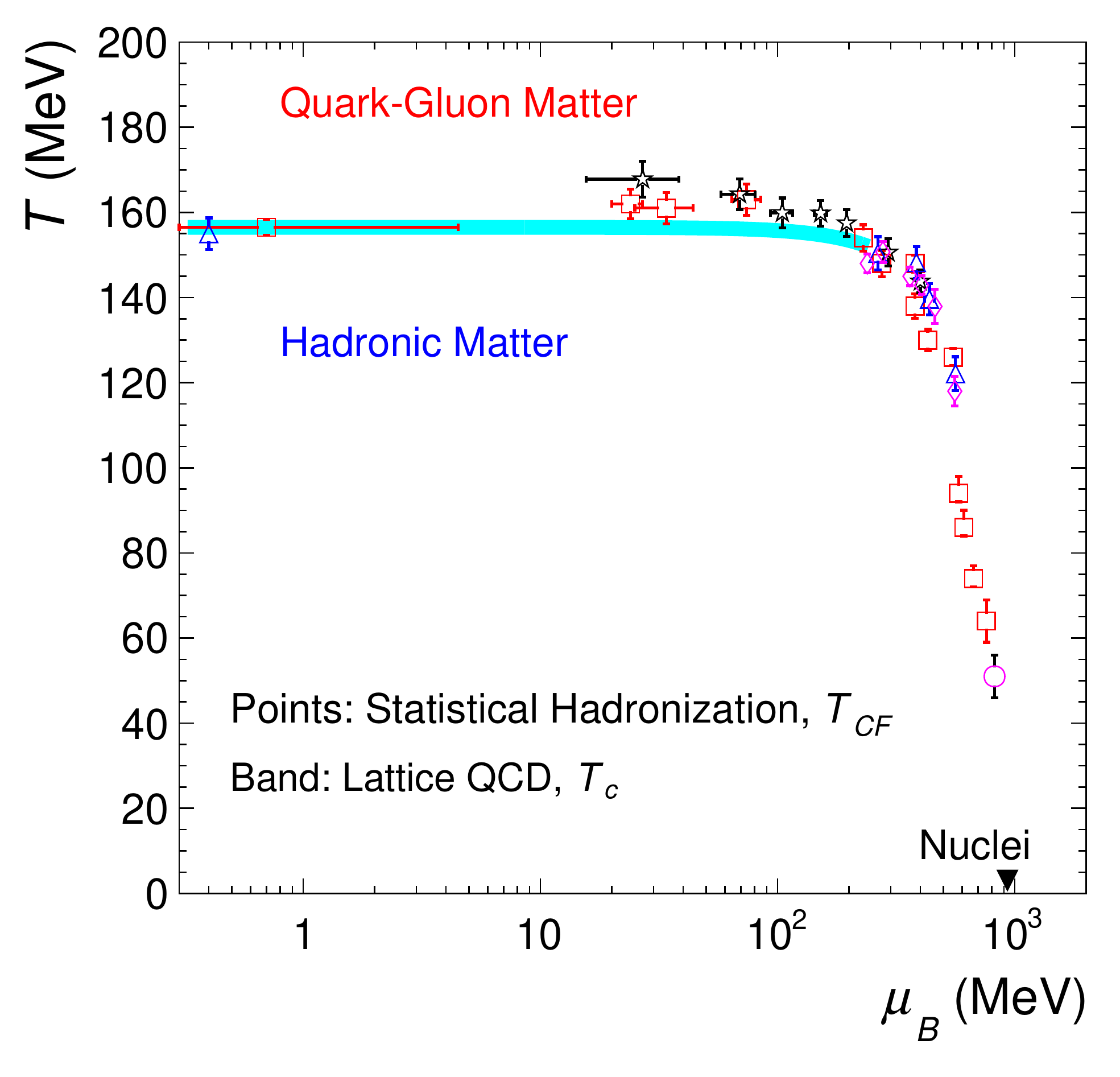}
\end{minipage} & \begin{minipage}{.47\textwidth}
\caption{Phenomenological phase diagram of strongly interacting matter constructed from chemical freeze-out points for central collisions at different energies, extracted from experimental data sets in our own analysis (squares) and other similar analyses \cite{Cleymans:1998yb,Vovchenko:2015idt,Becattini:2016xct,Adamczyk:2017iwn} are compared to predictions from LQCD \cite{Bazavov:2018mes,Borsanyi:2020fev} shown as a band.  The inverted triangle marks the value for ground state nuclear matter (atomic nuclei).}
\label{fig:t-mu}
\end{minipage}\end{tabular}
\end{figure}

Since the statistical hadronization analysis at each collision energy yields a pair of ($T_{CF}$,\mub) values, these points can be used to construct a $T$ vs. \mub diagram, shown in Fig.~\ref{fig:t-mu}. Note that the points at low temperature seem to converge towards the value for ground state nuclear matter ($\mub = 931$ MeV). 
As argued in \cite{Floerchinger:2012xd} this limit is not necessarily connected to a phase transition.  While the situation at low temperatures and collision energies is complex and at present cannot be investigated with first-principle calculations, the high temperature, high collision energy limit allows a quantitative interpretation in terms of fundamental QCD predictions.

\section{Statistical hadronization of charm quarks}  
An interesting question is whether the production of hadrons with heavy quarks can be described with similar statistical hadronization concepts.
We note that the mass of the charm quark, $m_c \simeq 1.3$ GeV, is sufficiently larger than the pseudo-critical temperature $T_c$ introduced above, such that thermal production of charm quarks is strongly Boltzmann suppressed, and that at the LHC a copious production of charm quarks in relativistic nuclear collisions through hard scattering processes is expected.
The produced charm quarks will, therefore, not resemble a chemical equilibrium population for the temperature $T$. However, what is needed for the thermal description proposed is that the heavy quarks produced in the collision reach a sufficient degree of thermal equilibrium through scattering with the partons of the hot medium. Indeed, the energy loss suffered by energetic heavy quarks in the QGP is indicative of their ``strong coupling'' with the medium, dominated by light quarks and gluons.  
The measurements at the LHC \cite{ALICE:2012ab,Abelev:2013lca} and  RHIC \cite{Adamczyk:2014uip} of the energy loss and hydrodynamic flow of  D mesons  demonstrate this quantitatively.

Among the various suggested probes of deconfinement, charmonium (the bound states of $c\bar{c}$) plays a distinctive role. The \jpsi meson is the first hadron for which a clear mechanism of suppression (melting) in the QGP was proposed early on, based on the color analogue of Debye screening \cite{Matsui:1986dk}.  
A novel quarkonium production mechanism, based on statistical hadronization was proposed \cite{BraunMunzinger:2000px}, based on thermalized charm quarks which are "distributed" into hadrons at the phase boundary, i.e. at chemical
freeze-out, with thermal weights as discussed above for the light
quarks, \cite{Andronic:2006ky,BraunMunzinger:2000px,BraunMunzinger:2009ih,Andronic:2011yq}.  An alternative mechanism for the (re)combination of charm and anti-charm quarks into charmonium in a QGP \cite{Thews:2000rj} was proposed based on kinetic theory (for further developments see \cite{Zhao:2011cv,Zhou:2014kka}).

In the SHM, the absence of chemical equilibrium for heavy quarks is accounted for by introducing a fugacity $g_c$. The parameter $g_c$ is obtained from the balance equation \cite{BraunMunzinger:2000px} which accounts for the distribution of all initially produced heavy quarks into hadrons at the phase boundary, with a thermal weight constrained by exact charm conservation. With the above approach the knowledge of the heavy quark production cross section along with the thermal parameters obtained from the analysis of the yields of hadrons composed of light quarks, see previous section, is sufficient to determine the yield of hadrons containing heavy quarks in ultra-relativistic nuclear collisions.

\begin{figure}[hbt]
\begin{tabular}{cc}  \begin{minipage}{.47\textwidth}
\includegraphics[width=1.0\textwidth]{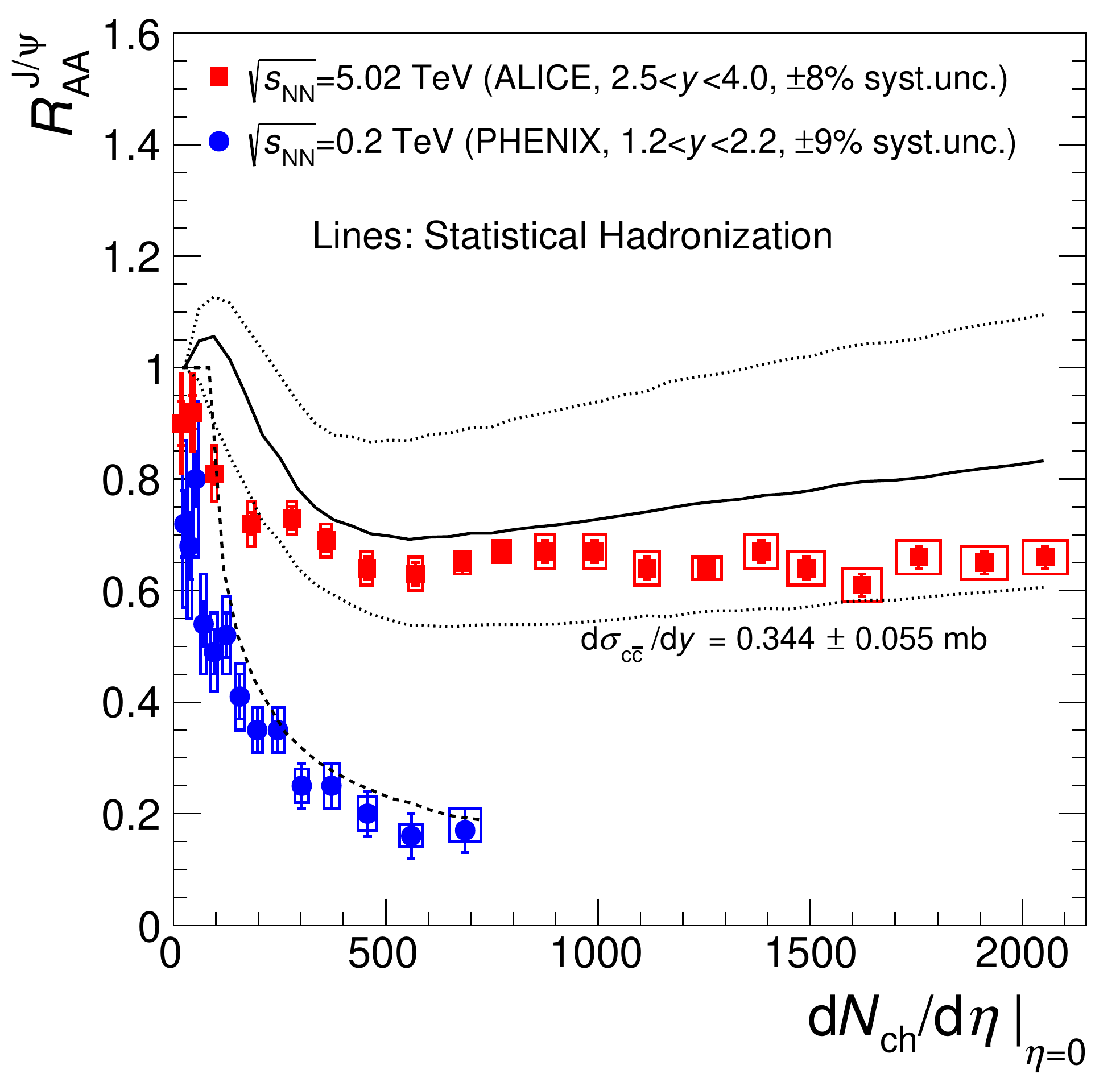}
  \end{minipage} & \begin{minipage}{.47\textwidth}
\includegraphics[width=1.0\textwidth]{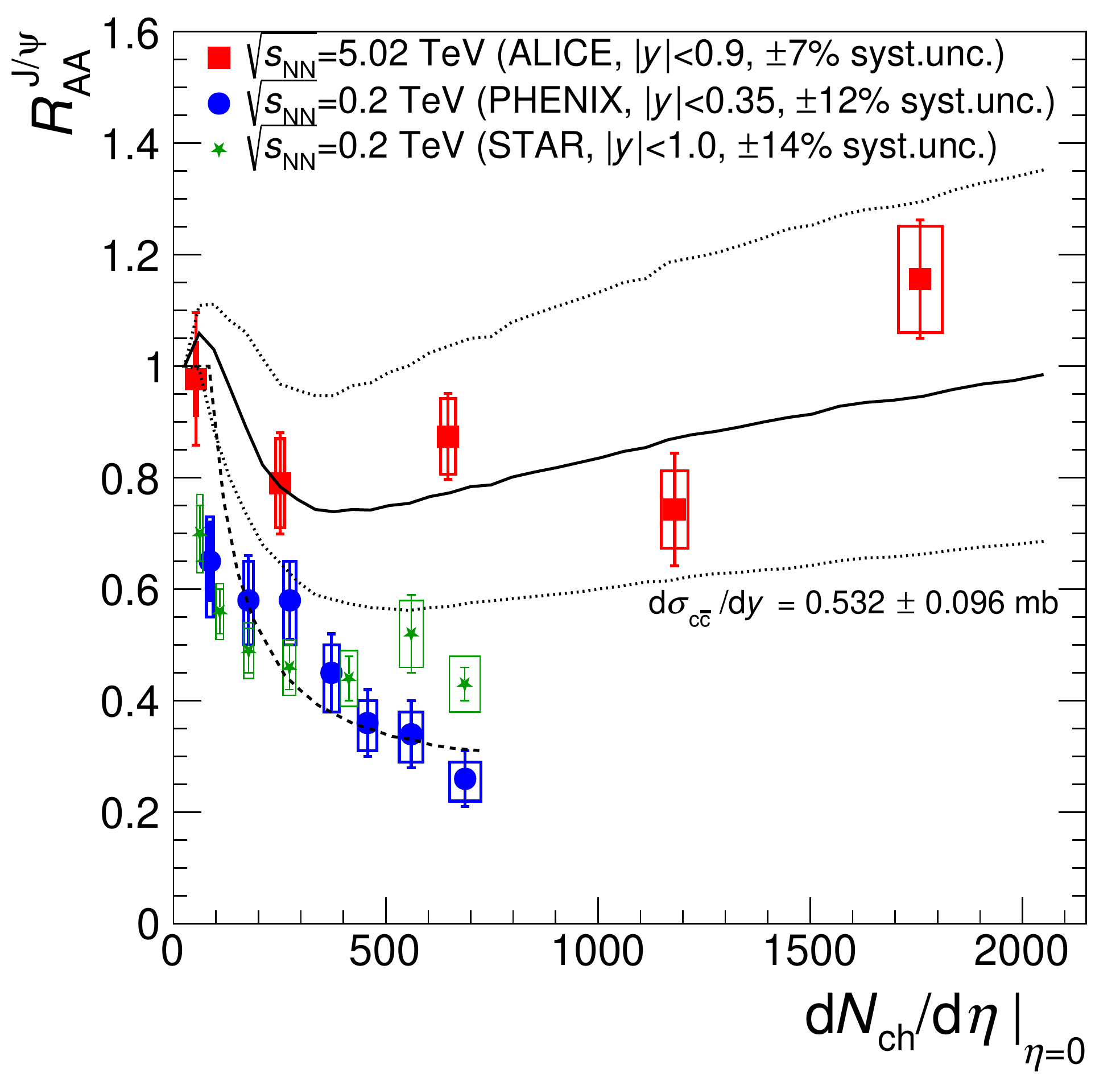}
\end{minipage}\end{tabular}
\caption{The nuclear modification factor $R_{\mathrm{AA}}$ for inclusive \jpsi production in dependence on the multiplicity density (at $\eta$=0) at forward rapidity (left panel) and at midrapidity (right panel). The data are for Au--Au collisions from PHENIX (blue) \cite{Adare:2006ns,Adare:2011yf} and STAR  (green) \cite{Adamczyk:2013tvk} at RHIC and for Pb--Pb collisions from the ALICE collaboration (red) \cite{Abelev:2013ila,Acharya:2019lkh} at the LHC.} \label{fig:raa_jpsi1}
\end{figure}

In the SHM, the \jpsi nuclear modification factor $R_{\mathrm{AA}}$ is obtained by computing the yields in AA collisions while the yields in proton-proton collisions are taken from experimental  data.  The so determined $R_{\mathrm{AA}}$ is predicted to increase with increasing collision energy \cite{Andronic:2007bi}, implying reduced suppression or even enhancement due to the rapid increase with energy of the charm production cross section.  Clear evidence for such a pattern was obtained with the first ALICE measurements at LHC energy \cite{Abelev:2013ila}. Since then a large number of additional data including detailed energy, rapidity, centrality and transverse momentum dependences of $R_{\mathrm{AA}}$ for \jpsi as well as hydrodynamic flow \cite{Acharya:2020jil} 
results have provided a firm basis for the statistical hadronization scenario \cite{Andronic:2006ky}, with the biggest uncertainties still related to the not yet measured value of the open charm cross section in Pb--Pb collisions. Current results on \jpsi production at midrapidity and forward rapidity as a function of the charged particle multiplicity and description within the SHM  are summarized in Fig.~\ref{fig:raa_jpsi1}. A dramatic increase of $R_{\mathrm{AA}}$ with increasing collision energy is clearly observed. Furthermore, the measurements at the LHC demonstrate \cite{Abelev:2013ila,Acharya:2019lkh}, that the increase is largely concentrated at \jpsi transverse momentum values less than the mass $m_{\jpsi} = 3.1$ GeV. This observation, first predicted within a transport model approach in Refs.~\cite{Zhao:2011cv}, is a natural feature for the statistical hadronization approach \cite{Andronic:2019wva}. 
The success of the SHM in the charm sector which provides a natural explanation of the increase with collision energy of $R_{AA}$ for J/$\psi$ is deeply connected to and provides unique evidence for the deconfinement of charm quarks \cite{Andronic:2017pug,BraunMunzinger:2000px} in the hot medium.

\vspace{.5cm}
\textbf{Acknowledgements}
K.R. acknowledges partial support by the Extreme Matter Institute EMMI
and the Polish National Science Center NCN under Opus grant no.2018/31/B/ST2/01663. This work is part of and supported by the DFG Collaborative Research Center ''SFB1225/ISOQUANT''.

\bibliography{wroclaw2020}

\end{document}